\documentclass[]{AO4ELT}  

\usepackage{microtype}
\usepackage[style=numeric,sorting=none]{biblatex} 
\usepackage{amsmath,amsfonts,amssymb}
\usepackage{graphicx}
\usepackage{pst-all} 
\usepackage[colorlinks=true, allcolors=blue]{hyperref}
\usepackage{listings}
\usepackage{tabularx}
\usepackage[section]{placeins}

\makeatletter         
\def\@maketitle{
\includegraphics[width = 170mm]{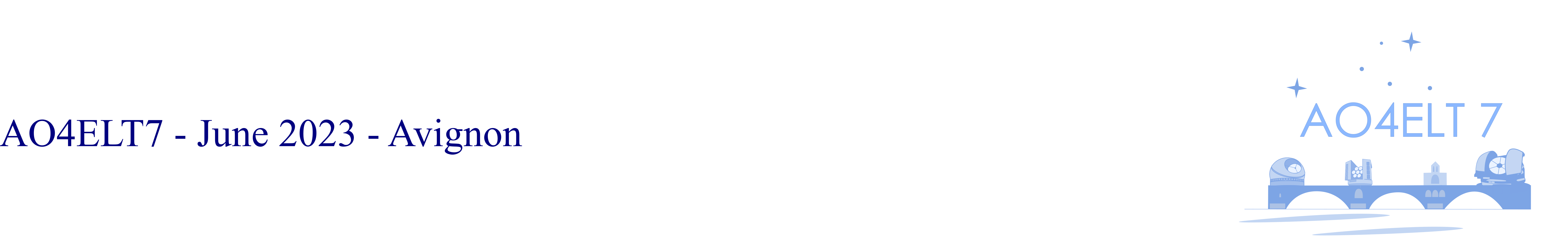}\\[8ex]
\begin{center}
{\Huge \bfseries \sffamily \@title }\\[4ex] 
{\Large  \@author}\\[4ex] 
\@date
\end{center}}

\title{Simulating METIS' SCAO System}

\author[a]{Markus Feldt}
\author[a]{Horst Steuer}
\author[a,b]{Carlos Correia}
\author[c]{Andreas Obereder}
\author[c]{Stefan Raffetseder}
\author[a]{Thomas Bertram}
\author[c]{Julia Shatokina}
\author[d]{Faustine Cantalloube}

\affil[a]{Max Planck Institute for Astronomy, K\"onigstuhl 17, D-69117 Heidelberg, Germany}
\affil[b]{Space ODT - Optical Deblurring Technologies Ltd., Porto, Portugal}
\affil[c]{MathConsult GMBH, Altenbergerstraße 69, A-4040 Linz, Austria}
\affil[d]{Laboratoire d'Astrophysique de Marseille, 38 Rue Frédéric Joliot Curie, 13013 Marseille, France}

\authorinfo{Further author information: (Send correspondence to M.F.)\\M.F.: E-mail: mfeldt@mpia.de, Telephone: +49 6221 528 262}

\begin{document} 
\maketitle

\begin{abstract}
METIS, the Mid-Infrared ELT Imager and Spectrograph, is one of the four first-generation ELT instruments scheduled to see first light in 2028. Its two main science modules are supported by an adaptive optics system featuring a pyramid sensor with 90x90 subapertures working in the $H$ and $K$ bands. During the PDR and FDR phases, extensive simulations were carried out to support the sensing, reconstruction, and control concept of METIS single-conjugate adaptive optics (SCAO) system.

We present details on the implementation of the COMPASS-based environment used for the simulations, the metrics used for analyzing our performance expectations, an overview of the main results, and some details on special cases like non-common path aberrations (NCPA) and water vapor seeing, as well as the low-wind effect.
\end{abstract}


\section{METIS SCAO}
\subsection{Overview}
\begin{figure}
\centering
\includegraphics[width=\textwidth]{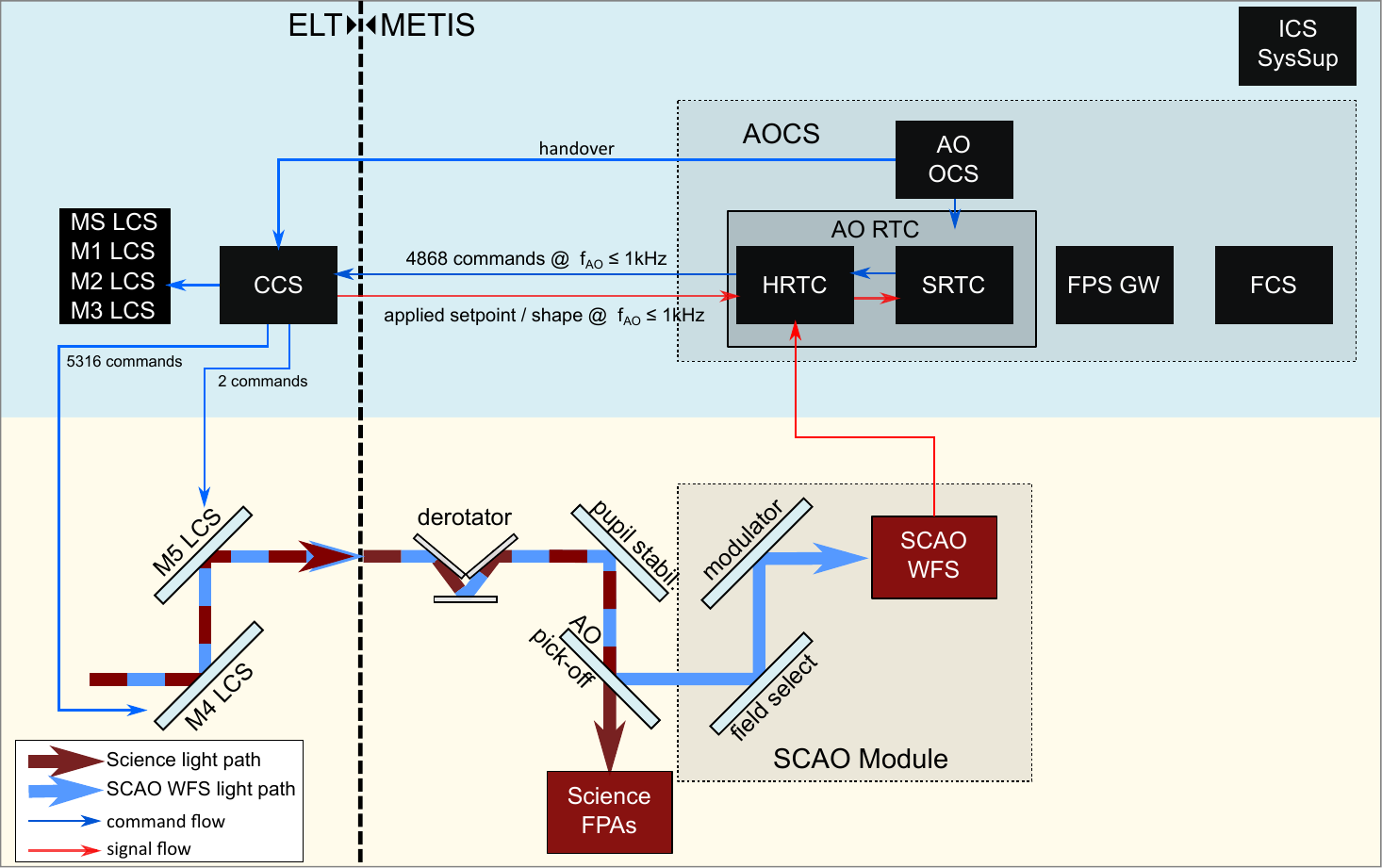}
\caption{Simplified block diagram of the SCAO system: The Adaptive Optics Control System (AOCS) and the SCAO Module (slightly darker boxes) are the entities of the SCAO system that belong to the instrument domain. The key entities for
the real-time correction of the incoming light are located in the 'ELT' domain on the left side of the figure. In a closed wavefront control loop, the blue, near-infrared (NIR) light is used to measure the instantaneous residual
wavefront error by the WFS. The measurement signal is analyzed by the RTC, and a computed correction is sent to the Central Control System (CCS) to be applied with the M4 and ELT tip-tilt field stabilization mirror (M5) via a Local
Control System (LCS). The Focal Plane Sensor Gateway (FPS GW) provides science images to auxiliary AO loops.}
\label{fig:SCAO-overview}
\end{figure}

Figure \ref{fig:SCAO-overview} depicts schematically the METIS SCAO system and its main submodules. The core of the system is the pyramid wavefront sensor (PWFS), using 90$\times$90 subapertures for sensing in the $K$ and $H$ bands, operating at a loop frequency of 1\,kHz. Fig.~\ref{fig:SCAO-overview} shows that in the common path in front of the SCAO module, there is a derotator unit and a pupil stabilization mirror. When the derotator is commanded to stabilize the field for science operations, the pupil image on the PWFS will rotate. The pupil stabilization mirror is controlled by SCAO to maintain registration between the SCAO unit and the M4 actuator grid, and/or the alignment of the coronagraphic train. Additionally, an apodizer mask causing interesting complications can be inserted into the the common optical for the apodized vortex coronagraph (AVC). More information on the METIS SCAO system can be found in \cite{bertram23b}.

\subsection{Wavefront Control Strategy}

\begin{figure}[htb]
\centering
\includegraphics[width=0.5\textwidth]{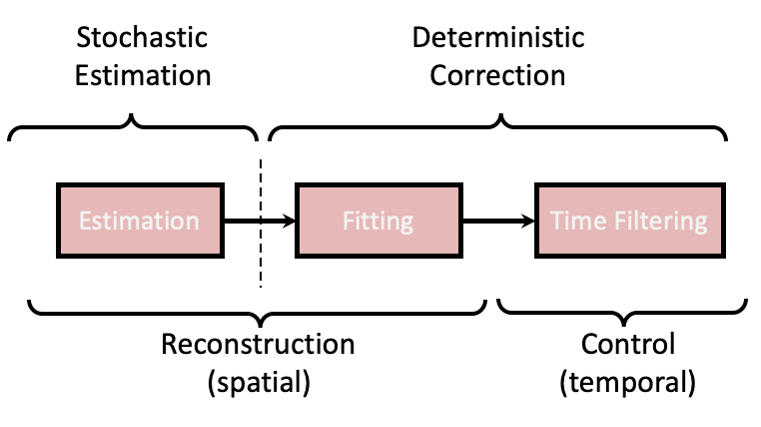}
\caption{Separation between stochastic estimation and deterministic correction, spatial reconstruction and temporal, dynamic control. This oversimplification shows the interplay of the different, spatial and temporal, stochastic and deterministic operations. The estimation and fitting are two distinct steps. This is a feature we retained since the pupil and DM actuator mesh optically rotate in front of the PWFS. Decoupling the reconstruction from the fitting is an aspect of significance that is considered in our developments for it allows for a streamlined implementation and AO system adaptation to changing conditions.}
\label{fig:WFC-Strategy}
\end{figure}

The METIS baseline control strategy was put together with the following goals in mind:

\begin{enumerate}
\item Controlling the maximum number of M4 degrees-of-freedom without generating unwanted spurious signals (e.g. differential-piston modes)
\item Performing numerical derotation and streamlined on-the-fly reconstructor updates, including mis-registration management
\item Using a set of aperture-spanning modes as a control basis ranked jointly by increasing spatial frequency and force on M4
\end{enumerate}

Based on these, we have converged on a split spatial reconstruction/temporal filtering, which is schematically represented in Fig.~\ref{fig:WFC-Strategy}. In our solution, we first solve for the wave-front spatial reconstruction neglecting any wavefront time evolution and postulate an a-priori Proportional-Integral (PI) structure for the time-filtering (a.k.a. regulators or controllers), both for Tip-Tilt (TT) and High Order (HO) modes. The reconstruction step is further sub-divided into two steps: i) the stochastic wavefront estimation from P-WFS data and ii) the deterministic fitting/projection onto the desired control space. The strategy outlined here tackles both steady-state and transient behavior.

In a nutshell, our wavefront control strategy looks as follows:

\begin{itemize}
\item Spatial reconstruction
\begin{itemize}
\item Tikhonov-regularised stochastic estimation expressed on the space formed by the influence functions of a Cartesian-mesh virtual DM layout
\item A fitting/projection step onto control modes
\begin{itemize}
\item Force-aware Karhunen-Loeve modes
\item Derotation and mis-registration compensation via linear algebra operations
\end{itemize}
\end{itemize}
\item Time filtering
\begin{itemize}
\item PI control for HO and TT modes
\item Anti-windup through error deflation using the CCS feedback (a.k.a. "echo")
\end{itemize}
\item Saturation avoidance
\begin{itemize}
\item Modal loop gain adjustment
\item Regularization strength adjustment on the fitting step between reconstruction and M4 space
\end{itemize}

\end{itemize}

\section{SIMULATION SETUP}

\begin{lstlisting}[frame=single, float=htb,language=python,caption={Standard parameters},label={lst:standpar}]
if not 'seeingCondition' in vars():
seeingCondition = 'median' # 'median', 'Q1', 'Q2','Q3','Q4'
if not 'zenithAngleInDeg' in vars():
zenithAngleInDeg = 30.

p_loop = conf.Param_loop()
p_loop.set_niter(60000) # Number of iterations
p_loop.set_ittime(0.001) # Loop frequency [s] 1000Hz

p_tel = conf.Param_tel()
p_tel.set_type_ap("EELT")
p_tel.set_diam(40.0) # Some obstruction already inbuilt
p_tel.set_t_spiders(0.54) # Spider size in meters

p_target = conf.Param_target()
p_target.set_Lambda(3.7) # Science wavelength = L-band
p_target.set_mag(4.537) # Star mag at the WFS wavelength
\end{lstlisting}

\subsection{COMPASS Modules \label{subsec:compass_modules}}
Simulations were carried out using the COMPASS environment \cite{github_compass} in version 5.0, a tool actively under development at LESIA \cite{2016SPIE.9909E..71G}. COMPASS has been extended by additional modules called {\tt p\_metis}, {\tt p\_calibration\_DM}, and {\tt p\_ccs}.

The first two modules essentially implement our reconstruction scheme using the virtual Fried-geometry DM and the regularized MMSE (Minimum Mean Squared Error) inversion and projection steps for the full set of available modes on M4 \cite{obereder23}. The {\tt p\_metis} module also includes features for operating on extended sources, applying NCPAs, dealing with non-ideal pyramid prism shapes, dynamic M1 segment aberrations, and more.

The {\tt p\_ccs} module takes the generated modal command vector and converts it into an M4 mirror shape, performing saturation avoidance checks as described in \cite{ESO-311982-V3}. It enforces limits on the maximum actuator gap (excursion from rest position), maximum speed, and maximum force exerted by each actuator. It also provides a corresponding echo signal to the controller.  The presence of the CCS module enables us to verify if our various robustness analyses also hold in the presence of limited available stroke and incompletely applied command requests.

\subsection{Baseline Configuration}

\begin{lstlisting}[frame=single, float=htb,language=python,caption={Reconstructor parameters},label={lst:recpars}]

Virtual DM for the reconstruction
p_calibration_DM = conf.Param_dm()
p_calibration_DM.set_type("pzt")
p_calibration_DM.set_nact(p_wfs0.nxsub + 1)
p_calibration_DM.set_alt(600.)
p_calibration_DM.set_unitpervolt(1.0)

p_calibration_DM.influ_type = 'pyramid' # FEM ansatz functions
p_calibration_DM.set_push4imat(0.05) # Stay in the linear regime of the PWFS
p_calibration_DM.set_coupling(0.0) # Allow proper FEM discretization
p_calibration_DM.set_thresh(0.03) # Don't cut off too strongly

p_metis = conf.Param_metis()
p_metis.set_flag_reconstructor_without_calibration_DM(False) 
p_metis.set_reconstructor_calibration_DM(p_calibration_DM)
p_metis.set_reconstructor_alpha_factor(0.1) # 0.1: conservative choice

p_metis.set_virtual_dm_actuators_filter_method('useNormThresholding')
p_metis.set_nullflatslopes(True) # Necessary when using norm thresholding
p_metis.set_use_modal_control_basis(True)

p_metis.set_wavefront_fitting_method('ModalViaM4')
p_metis.set_wavefront_fitting_alpha_factor(0.1) # Conservative choice, VERY stable
p_metis.set_fitting_regularisation_matrix_filename('foobar.npy')

p_metis.set_modes_file('foobar2.npy')
p_metis.set_modes_use_full_basis(True)
p_metis.set_number_controlled_modes(ALL)
\end{lstlisting}

Throughout the different phases of the project, we used so-called baseline configurations, and deviations from these baselines were made only for specific analyses in small subsets of the parameter set. The main baseline configuration for phase D, the post-FDR phase of METIS, is shown in Listing \ref{lst:standpar}. It specifies a bright star in median atmospheric conditions at 30 degrees zenith distance seen through an ELT pupil with 54 cm spiders. Our standard simulation runs for 60 seconds at 1 kHz.

Listing \ref{lst:recpars} shows the setup of the reconstructor using a virtual deformable mirror (VDM) tied to the PWFS's pixel grid in a Fried geometry. The VDM uses bilinear ansatzfunctions as influence functions ("pyramidlets"), something a physical DM never could. The virtual calibration is done with an actuator push amplitude (aka {\tt push4imat}) of 50 nm to stay well within the sensor's linear regime. The reconstructed wavefront is projected onto a set of force-aware Karhunen-Loeve modes \cite{correia23}. We always project onto and control the full set of modes, with higher-order modes receiving sequentially smaller amplitudes via the regularization mechanism \cite{obereder23}.

Listing \ref{lst:cntpars} describes the setup of the PI controller for high-order and tip-tilt control. The {\tt P\_gain} and {\tt I\_gain} factors $gI$ and $gP$ translate into control commands according to:

\begin{equation}
\mathbf{m}_k = \gamma \mathbf{m}_{k-1} + (gI + gP)\mathbf{e}_k - gP \mathbf{e}_{k-1},
\end{equation}

where $\mathbf{m_k}$ denotes a modal command at loop step $k$, $k$ is the loop index, and $\mathbf{e_k}$ is the error signal at loop step $k$. $\gamma$ is the leakage factor with $0 < \gamma \leq 1$ in theory, but $\gamma$ is very close to 1 in practice.

\begin{lstlisting}[frame=single, float=htb,language=python,caption={Controller parameters},label={lst:cntpars}]
p_controller0 = conf.Param_controller()
p_controller0.set_type("metis")
p_controller0.set_delay(1.1) # PARTIAL DELAYS in units of the integration time T.
p_controller0.set_gain(1 / (1 + p_controller0.get_delay()))

p_metis.set_ctrl_tt_P_gain(0.3)
p_metis.set_ctrl_tt_I_gain(0.29)
p_metis.set_ctrl_hm_P_gain(0)
p_metis.set_ctrl_hm_I_gain(0.45)
\end{lstlisting}

Listing \ref{lst:ccspars} shows the setup of the CCS module according to \cite{ESO-311982-V3}. The module enforces limits on the maximum actuator gap (30 $\mu$m), maximum speed (2.5 m/s), and maximum force exerted by each actuator (1.2 N). 

CCS enforces these limits in a dedicated scheme:

\begin{enumerate}
    \item Calculate a DM shape from the first 50 modes, and check for gap (30\,$\mu$m) and speed (2.5\,m/s).  If either limit is violated, scale the modal command vector in order to scale the requested wavefront accordingly.
    \item Check the maximum applied actuator force. If more than 1.2\,N are requested from any actuator, cut the highest 40 modes from the command vector, and recalculate. Repeat until the maximum requested force is below 1.2\,N.
    \item Calculate gap and speed on the resulting requested DM shape. If either criterion is violated, drop the frame and re-apply the last command.
    \item Send an echo of the actual applied DM shape to the controller.
\end{enumerate}

\begin{lstlisting}[frame=single, float=htb,language=python,caption={CCS parameters},label={lst:ccspars}]
p_ccs = conf.Param_ccs()
p_ccs.set_max_actuator_gap(30.0)
p_ccs.set_max_actuator_speed(2.5)
p_ccs.set_max_force(1.2)
p_ccs.set_modes_to_cut_iteratively(40)  
p_ccs.set_modes_in_gap_analysis(50)  
\end{lstlisting}

\subsection{AOSAT - Key Performance Parameters}

\begin{figure}
\centering
\includegraphics[width=0.7\textwidth]{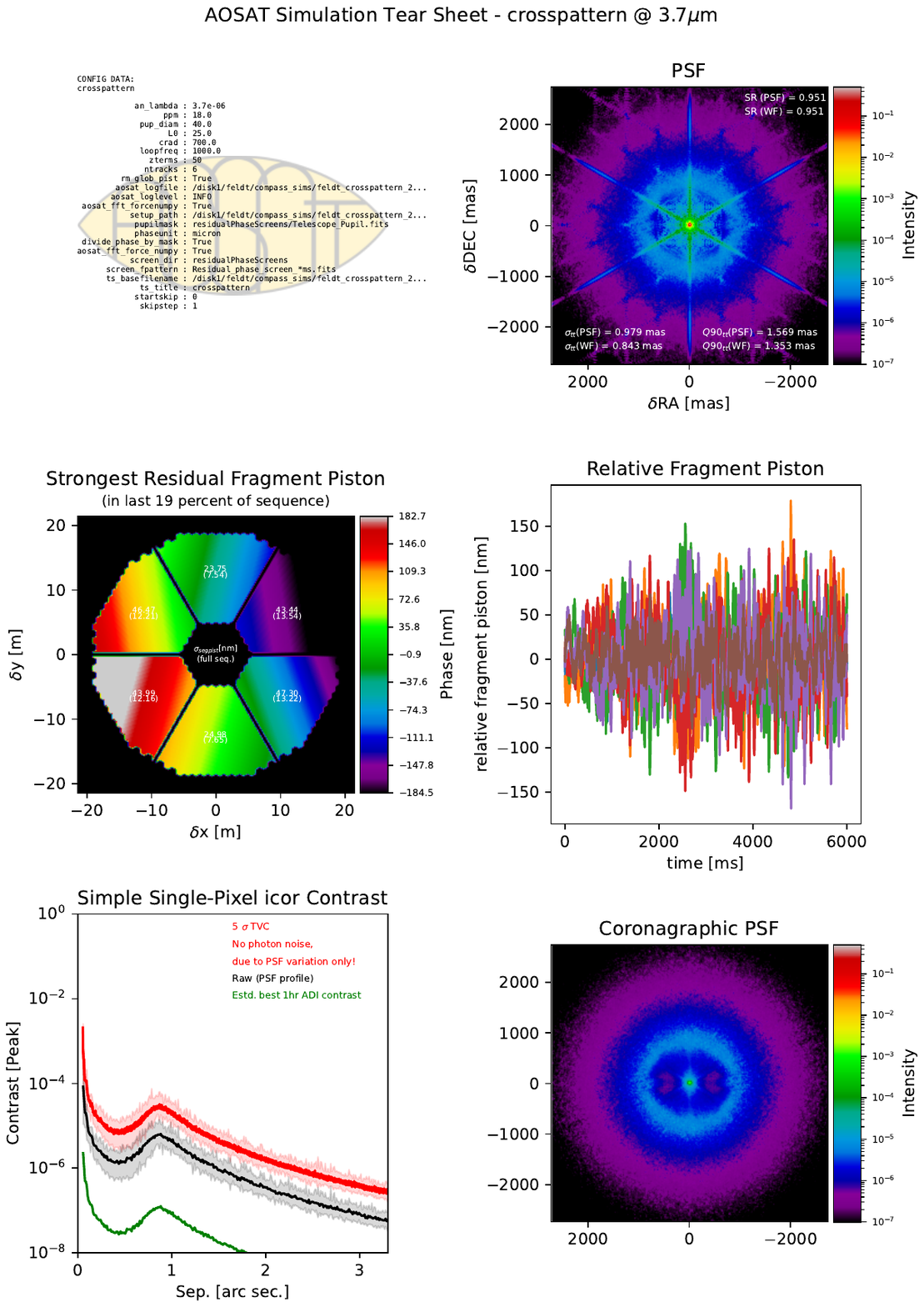}
\caption{Example output of the graphical summary sheet produced by AOSAT for a single simulation run. Simultaneously, a textual representation of the data is generated, which is used by the plotting scripts to produce graphs like the one shown in Fig.~\ref{fig:stdfig_example}.}
\label{fig:AOSAT_example}
\end{figure}

\begin{figure}
\centering
\includegraphics[width=0.7\textwidth]{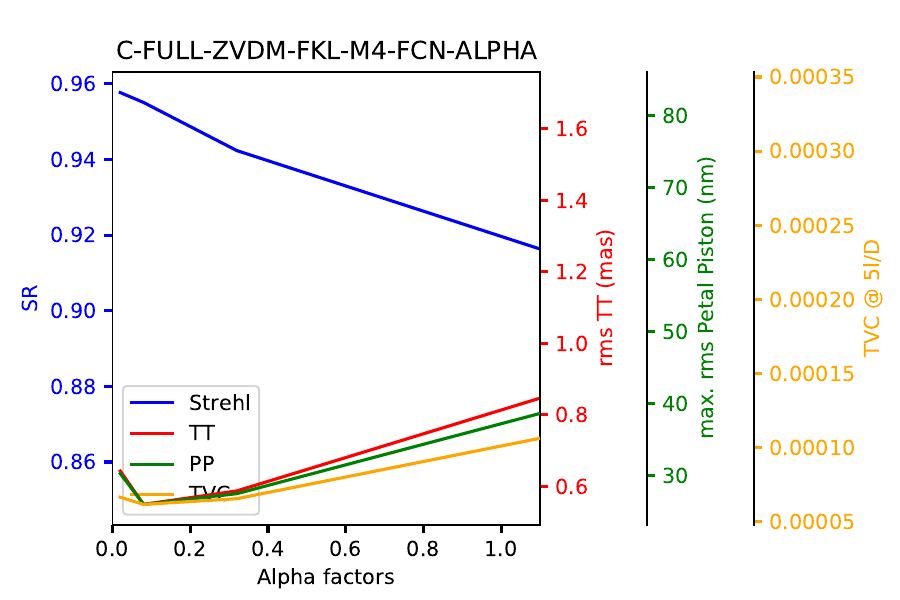}
\caption{Typical plot of our key performance parameters against a varying input parameter, in this case both regularization strengths of the reconstructor which were kept identical. The Strehl ratio is plotted in blue, residual tip-tilt motion of the PSF in red, residual petal piston in green, and the temporal variance contrast in orange.}
\label{fig:stdfig_example}
\end{figure}

The standard metric to watch when simulating adaptive optics is, of course, the Strehl ratio. However, in METIS, the requirements placed on Strehl are quite relaxed and are easily exceeded even in the worst conditions. The true challenging requirements are on the final contrast achieved and on residual image motion and petal piston. Therefore, we use the AOSAT\footnote{AOSAT can be obtained from github (github.com/mfeldt/AOSAT), the documentation is on readthedocs (aosat.readthedoccs.io).} package \cite{feldt20} for the analysis of SCAO simulation results. AOSAT provides a suite of "analyzers" for various performance metrics.

AOSAT is run after a given simulation has finished, utilizing the residual wavefront frames of each simulation step. The telescope pupil and the setup file required by AOSAT are automatically generated to match the circumstances of the simulation. AOSAT is GPU-accelerated, resulting in an analysis framerate between thrice and equal to that of COMPASS, depending on the analyzers in use.

The set-up used for most analyses includes analyzers for the PSF (Strehl ratio, residual TT jitter), the TVC (temporal variance contrast), pupil fragmentation (petal piston), and Zernike decomposition of the residual wavefronts. A summary sheet is saved in graphical and textual form, and the analyzer objects themselves are kept for more in-depth scrutiny of the recorded information. The residual phase screens are then deleted to save disk space.

AOSAT provides a range of analyzers, of which we use the following for our standard analyses:

\begin{itemize}
\item[PSF]
\begin{itemize}
\item Strehl from peak intensity
\item Strehl from average WF rms
\item Tip-tilt rms jitter from peak position
\item Tip-tilt rms jitter from WF
\item 90\%-quantile of tip-tilt excursion from peak position
\item 90\%-quantile of tip-tilt excursion from WF
\end{itemize}

\item[TVC] temporal variance contrast for ideal and no coronagraph
\begin{itemize}
\item 5$\sigma$ variance contrast mean at given separation
\item 5$\sigma$ variance contrast max at given separation
\item 5$\sigma$ variance min at given separation
\item raw contrast mean at given separation
\item raw contrast max at given separation
\item raw min at given separation
\end{itemize}

\item[PHS] phase analyzer
\begin{itemize}
\item Average rms of WF
\end{itemize}

\item[FRG] pupil fragmentation
\begin{itemize}
\item Fragment IDs (yielding e.g. the number of pupil fragments found)
\item Average piston of fragments
\item RMS piston of fragments
\item Average tip of fragments
\item RMS tip of fragments
\item Average tilt of fragments
\item RMS tilt of fragments
\end{itemize}

\item[ZRN] Zernike (or other modal) decomposition
\begin{itemize}
\item Mean of modal amplitudes
\item RMS of modal amplitudes
\end{itemize}

\end{itemize}

The fragment piston analyser provides both the differential piston mode on each segment (after the global mean over the pupil is removed) and the tilt-removed differential piston. In essence fragment piston is neither one nor the other -- but something in-between, since we cannot isolate the fragmentation portion out of the remainder modes. 

The TVC analyzer computes the temporal variance of the PSF in each pixel, making it a good proxy for the final post-processed contrast achieved by angular differential imaging methods in the absence of slow-varying aberrations such as NCPAs \cite{feldt20}.  An extensive analysis of METIS' high-contrast performance based on a full end-to-end simulation including the complete data reduction and analysis can be found in \cite{absil23}.

When analyzing the robustness of the system or optimizing its performance, we generally track four key performance parameters: The Strehl ratio at 3.7 $\mu$m, residual image motion, petal piston, and the TVC. This typically results in plots as shown in Fig.~\ref{fig:stdfig_example}.

\section{ROBUSTNESS ANALYSES}
\subsection{Baseline Performance}

\begin{figure}[htb]
\centering
\includegraphics[width=0.7\textwidth]{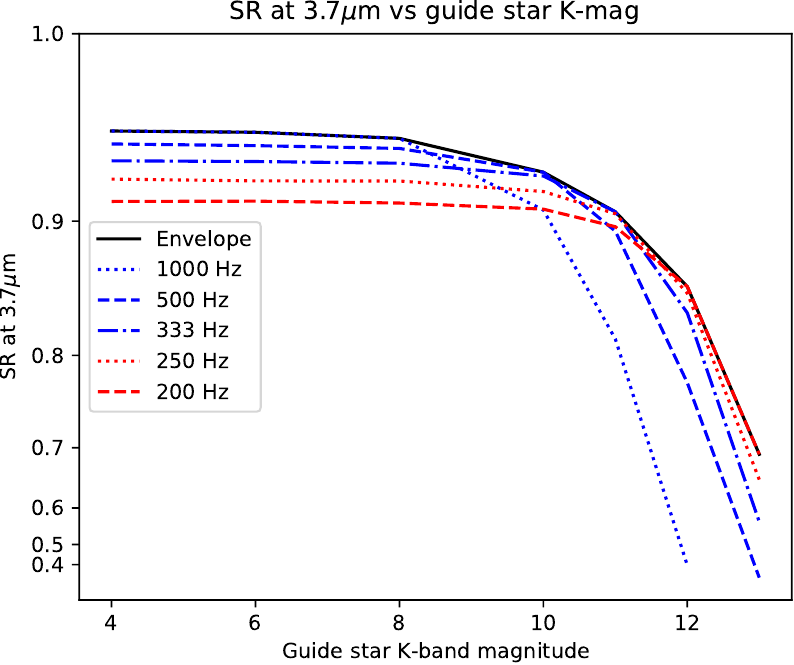}
\caption{Performance of METIS SCAO in post-FDR baseline configuration (medium seeing, 30$^\circ$ zenith distance, standard setup) versus guide star magnitude.}
\label{fig:baseline_performance}
\end{figure}

Figure \ref{fig:baseline_performance} depicts the expected performance degradation with increasing guide-star magnitudes (i.e. fainter stars). A Strehl ratio (at 3.7\,$\mu$m) above 90\% is maintained down to m$_K$=11,mag, and at m$_K$=13,mag, the end of the simulated sequence, we still fulfill the METIS Strehl requirement of achieving 60\% Strehl at 3.7\,$\mu$m. At that point, the loop is operating at 200\,Hz with an overall control gain of 0.3.

\subsection{NCPA}

\begin{figure}[htb]
\centering
\includegraphics[width=0.7\textwidth]{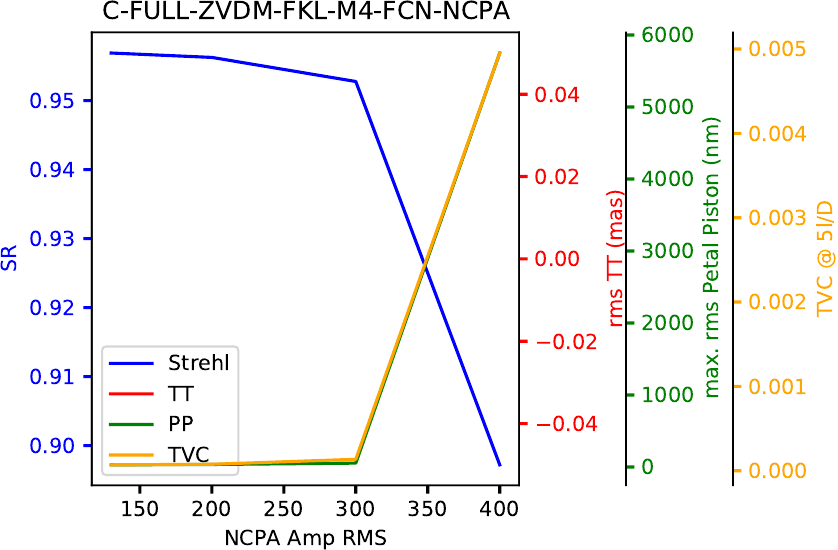}
\caption{SCAO performance as a function of applied NCPAs. A number of optimizations were applied, see text.}
\label{fig:C-FULL-ZVDM-FKL-M4-FCN-NCPA}
\end{figure}

In METIS, optical NCPAs were originally expected to be below 130\,nm, but current NCPA maps from the optical design only show values up to 92\,nm. Being conservative, we still use the 130\,nm threshold as the expected value for NCPAs.

In this experiment, however, we included residual water vapor (WV) seeing, effectively introducing NCPAs between WFS and science path which follow a Kolmogorov-like power spectrum. In order to achieve this, 100 Zernike terms were fitted to an open-loop atmospheric phase screen, and the result was used as optical NCPAs. The rms from water vapor seeing is expected to be of the order of 300\,nm, reaching 400\,nm under bad conditions. We thus simulated four NCPA rms values of 130 (the upper limit for expected NCPA purely from optics), 200, 300, and 400\,nm.

In addition, we included wind-induced tip-tilt according to ESO-supplied power spectra for a wind speed of 8\,m/s blowing from 20$^\circ$ off the pointing direction.

The system is not easy to stabilize under the presence of both NCPA and wind-induced tip-tilt. In order to find a stable performance, a number of setup parameters were allowed to vary, most notably the modulation amplitude, the regularization factors (which were both, for reconstruction and projection, kept identical), and the loop gain (in this particular experiment, the controller was a pure integrator).

\begin{table}[htb]
\centering
\caption[WV NCPA Optimized Parameters]{Reconstructor and Control Parameters optimized for each amount of WV-NCPAs. The numbers quoted produce the optimal Strehl ratio as measured on the PSF, the numbers in brackets produce the optimum TVC. The regularization parameters for reconstruction and fitting $\alpha_r = \alpha_f$ were identically applied for reconstruction and projection of the wavefront.}
\label{tab:C-FULL-ZVDM-FKL-M4-FCN-NCPA}
\begin{tabular}{|c|ccc|} \hline
WV NCPA & Modulation & Regularization & control gain \\
rms [nm] & amplitude [$\lambda$/D] & $\omega$ & g \\ \hline
130 & 4 (4) & 0.02 (0.08) & 0.4 (0.5) \\
200 & 4 (4) & 0.02 (0.08) & 0.6 (0.5) \\
300 & 6 (6) & 0.02 (0.08) & 0.3 (0.6) \\
400 & 7 (7) & 0.32 (0.32) & 0.4 (0.4) \\
\hline
\end{tabular}
\end{table}

Fig.~\ref{fig:C-FULL-ZVDM-FKL-M4-FCN-NCPA} shows that we can indeed tolerate NCPAs of up to 300\,nm without significant loss of performance. At 400\,nm we can still operate the system in a stable manner, achieving a Strehl of 90\% at 3.7\,$\mu$m. Despite the water vapor seeing amplitude is growing with wavelength, we may still expect very good performance at 10\,$\mu$m. The best parameters to operate at each NCPA amount found are summarized in Tab.~\ref{tab:C-FULL-ZVDM-FKL-M4-FCN-NCPA}.

\subsection{Mis-registration, Mis-rotation}

\begin{figure}[htb]
\centering
\includegraphics[width=0.65\textwidth]{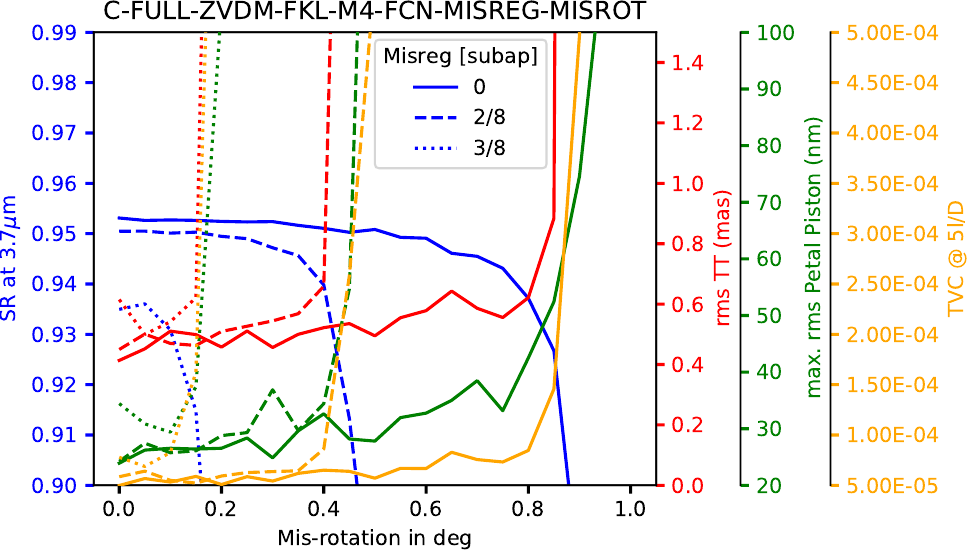}
\caption{Performance sensitivity to un-calibrated rotation with {0 (solid lines), 1/4 (dashed lines), and 3/8 (dotted lines)} of a PWFS sub-aperture (size=0.44\,m) offset.}
\label{fig:C-FULL-ZVDM-FKL-M4-FCN-MISREG}
\end{figure}

During closed-loop operation of METIS, the pupil will both drift and rotate. The same is true not only for the pupil but also for the image of M4, and thus the registration of its actuators to the PWFS. While we are keeping track of both effects with dedicated pupil and M4 registration monitoring algorithms and the pupil steering mirror, we may still experience conditions where the registration between M4 and the PWFS is different from what it was calibrated for.

We simulated such cases by calibrating the system in a nominal state and then offsetting and rotating the M4 model before commencing closed-loop operation. The result of the experiment is shown in Fig.~\ref{fig:C-FULL-ZVDM-FKL-M4-FCN-MISREG}. Due to numerical constraints in the simulation code, we can offset the M4 model in steps of an eighth of a pixel on the PWFS' detector. Our pupil tracking algorithm is specified to provide a pupil tracking accuracy of a tenth of a pixel on the PWFS, i.e. a tenth of a subaperture, and we expect the mis-registration identification scheme to provide a similar performance. We can thus assume that we can tolerate a pupil rotation of up to $\sim 0.5^\circ$ before a new command matrix (CM) needs to be uploaded to the hard real-time computer (HRTC).

\subsection{Low Wind Effect}

In conditions of very low wind speeds across the secondary support structure of any modern telescope, substantial temperature differences can occur between the air on one side of a girder and that on the other side. As a result, optical path differences of several microns between the two sides can be induced, difficult to sense and correct for any adaptive optics system. Frequently AO systems in such circumstances produce a petal pattern, where one fragment of the pupil delimited by support structures has a constant phase offset with respect to its neighbor, usually a multiple of the sensing wavelength.

\paragraph{Approximation by petal piston}

In early project phases, we assessed the resilience of our system with respect to such adverse effects by inducing said petal modes. In our case, sensing in the K-band, we do expect our system to be able to accommodate roughly half a sensing wavelength of wavefront discontinuity before the (differential-piston) error is conducted to the next integer multiple of $\lambda_{\text{wfs}}$. This is effectively true in the absence of turbulence. The latter, however, adds to the LWE modes at the beginning of loop closure. Despite it being continuous, the presence of naturally-occurring, relatively large differential-piston components may still trigger the residual error jumps to multiples of $\lambda_{\text{WFS}}$ despite the initial LWE mode being lower than the theoretical limit.

\begin{figure}[htb]
\centering
\includegraphics[width=0.45\textwidth]{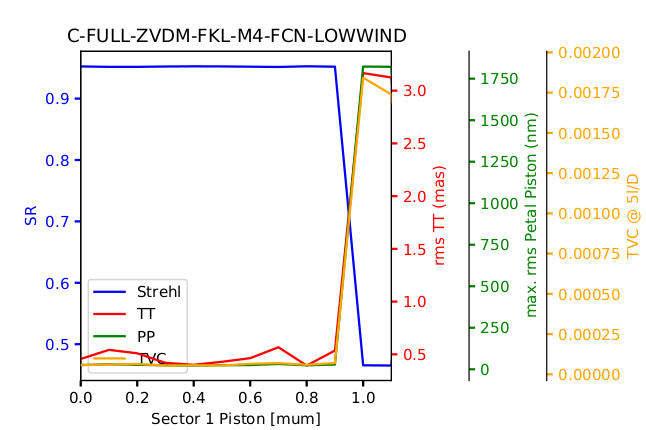}
\includegraphics[width=0.45\textwidth]{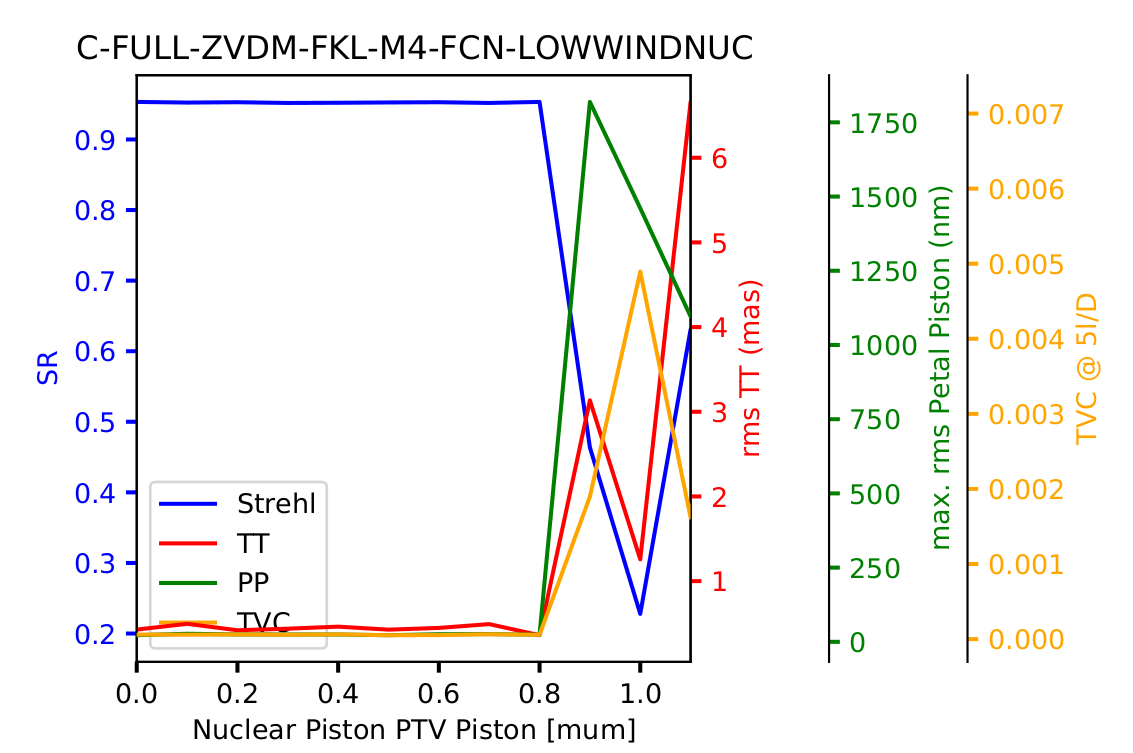}
\caption{Sensitivity as a function of initial amplitude of a single petal piston mode (left) and a vortex-shaped piston pattern (right).}
\label{fig:C-FULL-ZVDM-FKL-M4-FCN-PP}
\end{figure}

Figure \ref{fig:C-FULL-ZVDM-FKL-M4-FCN-PP} illustrates the sensitivity to a single petal piston introduced at loop closure as a function of its initial amplitude. As expected, for amplitudes lower than $\lambda_{\text{WFS}}/2$, the petal is correctly brought to the correct multiple of $\lambda_{\text{WFS}}$, whereas past this threshold the petal is wrongly conducted to the next multiple of $\lambda_{\text{WFS}}$. The exact value is below $\lambda_{\text{WFS}}/2$ on account of the uncorrected turbulence, which contains naturally-occurring piston that adds to the LWE at loop closure, making the system tip to the correct or incorrect multiple with an intrinsic variability.

The sensitivity results of Fig. \ref{fig:C-FULL-ZVDM-FKL-M4-FCN-PP} (right panel) show METIS is robust to around 3/4 $\lambda_{\text{WFS}}/2$, i.e. the presence of uncorrected atmospheric turbulence introduces a randomness factor on top of the LWE vortex mode that makes the petals tip to the wrong multiple of $\lambda_{\text{WFS}}$.

\paragraph{True optical path difference simulations}

\begin{figure}[htb]
\centering
\includegraphics[width=0.75\textwidth]{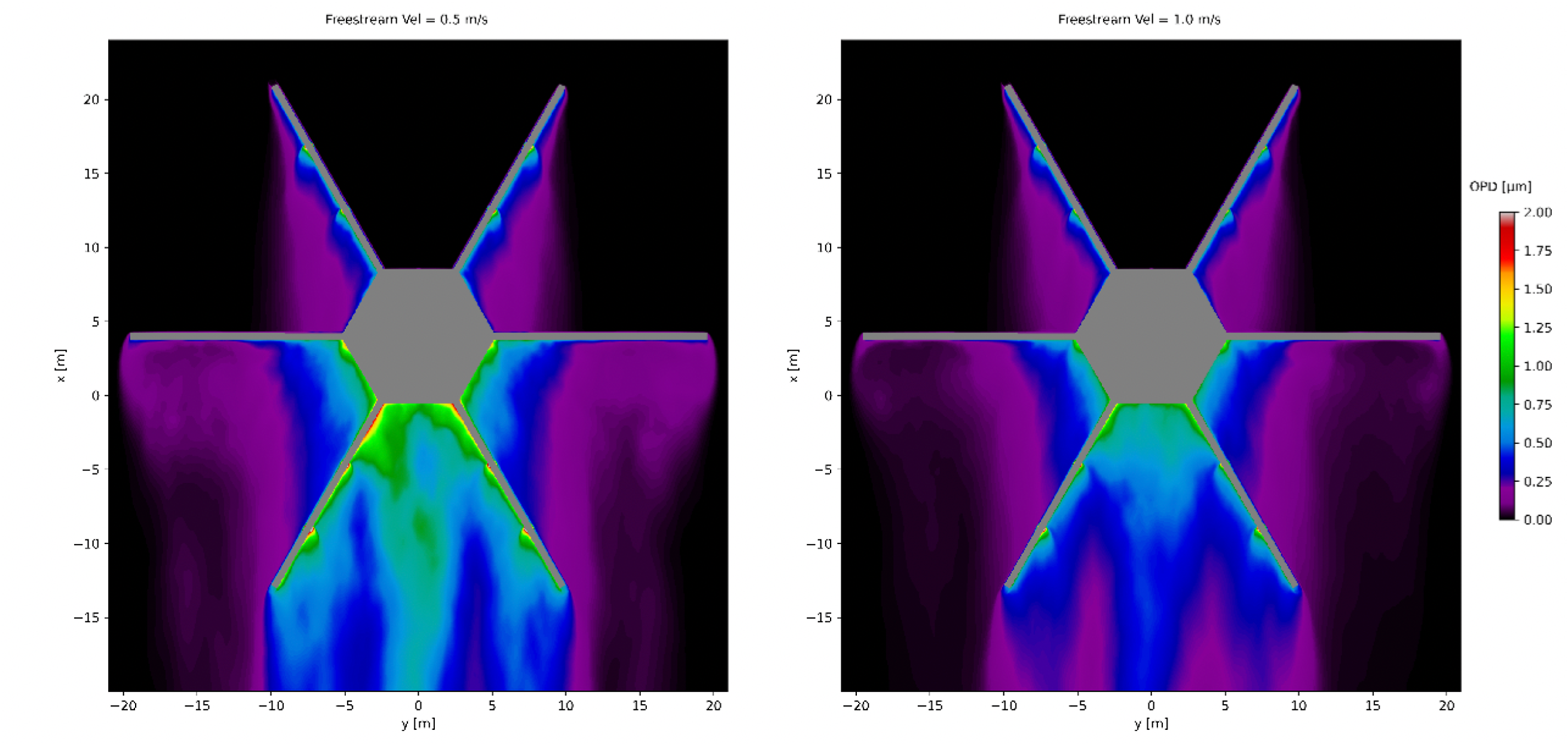}
\caption{Average OPD maps for 0.5\,m/s (left) and 1\,m/s (right) simulations. Extracted from Fig 8 of Martins et al (2022). Instantaneous OPD maps are provided within the paper.}
\label{fig:lweMapsFromMartins22}
\end{figure}

In addition to the early analysis provided above, we are now in possession of time-varying error maps obtained from computational fluid dynamic models for low-wind conditions \cite{martins22}. Such maps were kindly provided by Ronald Holzloehner and are shown in Fig.~\ref{fig:lweMapsFromMartins22} for two low-wind conditions.

\begin{figure}[htb]
\centering
\includegraphics[width=0.45\textwidth]{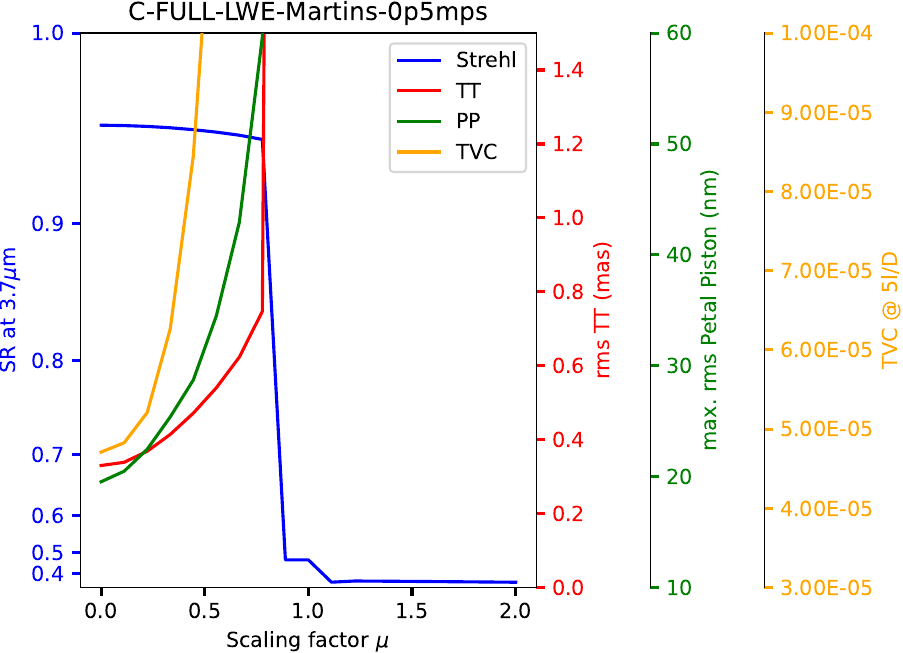}
\includegraphics[width=0.45\textwidth]{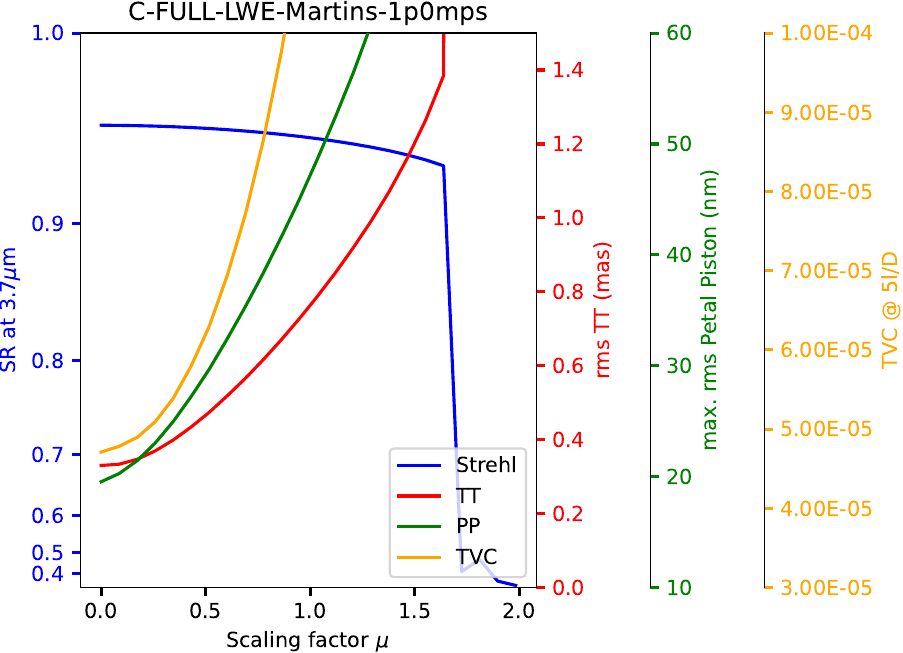}
\caption{Performance obtained in the presence of scaled low-wind-induced OPD maps for wind speeds of 0.5\,m/s (left) and 1.0\,m/s (right). The scaling factor $\mu$ is 1.0 where the nominal simulation results of \cite{martins22} are fully applied, including the scaling factor of 2.7 for the 0.5\,m/s and 1.74 for the 1.0\,m/s results.}
\label{fig:C-FULL-ZVDM-FKL-M4-FCN-LWE}
\end{figure}

The simulation setup was a the standard introduced after FDR. Deviating slightly from that, a higher regularization factor of 5.0 in the reconstructor during an initial loop closure phase of 1200 frames was applied. From experience, this helps to close the loop when the conditions are such that petalling is about to become problematic. Measurements of the standard parameter set as shown in Fig.~\ref{fig:C-FULL-ZVDM-FKL-M4-FCN-LWE} were performed on the 6000 frames following this closure period.

Keep in mind that "standard setup" implies a bright star and median seeing conditions; the latter might actually not necessarily apply during low-wind conditions.

In an additional experiment, we gradually introduced the LWE after the loop was closed, thus simulating a condition where the wind speed decreases gradually during normal operation.

Fig.~\ref{fig:C-FULL-ZVDM-FKL-M4-FCN-LWE} summarizes the performance metrics for the impact of the LWE in our standard configuration. It is obvious that without further mitigation, we cannot run the system under conditions as simulated by \cite{martins22} for 0.5\,m/s in median seeing when these conditions prevail already at loop start-up. The system immediately locks into a 2$\pi$ petalling state and remains so. For 1\,m/s, the situation is better, and we could cope with nominal results ($\mu$=1) in the sense of being able to close the loop and run in a stable fashion at least throughout our simulated 6s period. However, at $\mu$=1 there is already an impact on the performance metrics, most notably a contrast loss by a factor of $\sim$2.5.

\begin{figure}
  \begin{center}
    \includegraphics[width=0.7\textwidth]{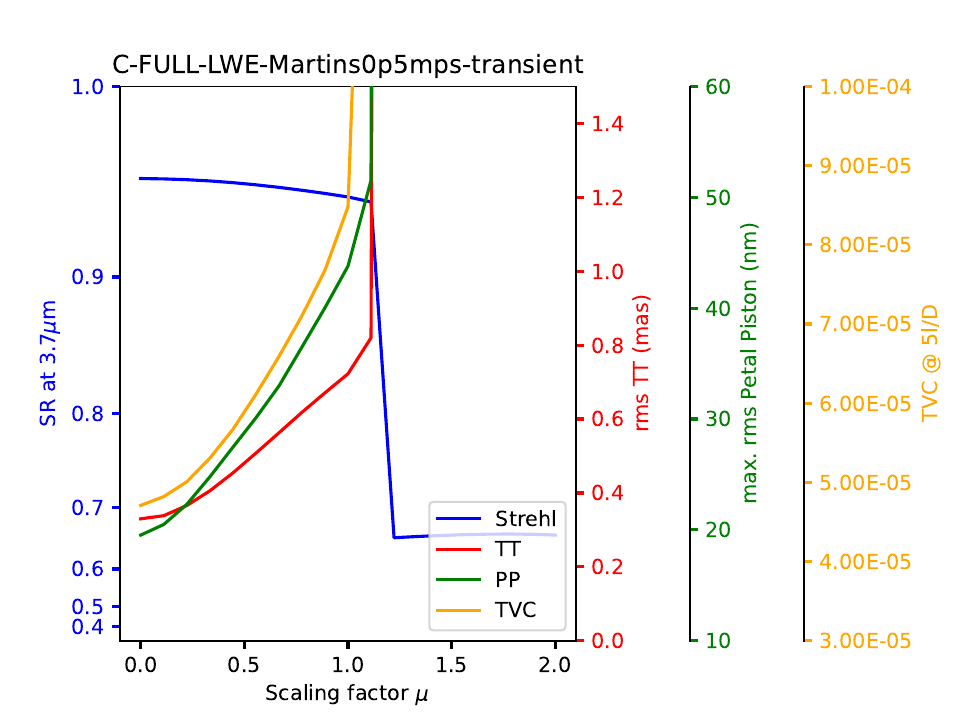}
  \end{center}
  \caption[]
  {\label{fig:C-FULL-ZVDM-FKL-M4-FCN-LWE-TRANSIENT}
    Performance obtained in the presence of scaled low-wind-induced OPD maps for wind speeds of 0.5\,m/s, gradually introduced over 0.2\,s after the loop had been closed. 
    }
\end{figure}

Fig.~\ref{fig:C-FULL-ZVDM-FKL-M4-FCN-LWE-TRANSIENT} summarizes our standard performance metrics for the case that the LWE is gradually introduced over 0.2\,s after the loop had been closed. In this case, we can cope with slightly larger amplitudes of the effect, and the loop remains stable at the nominal scaling factor ($\mu$=1) amplitude of the effect even at 0.5\,m/s. However, also here we do see a considerable impact on all performance metrics, again most notably a contrast loss by a factor of about $\sim$2.

\subsection{Safety First - Enforcing CCS Limits}

\begin{figure}
\begin{center}
\includegraphics[width=0.45\textwidth]{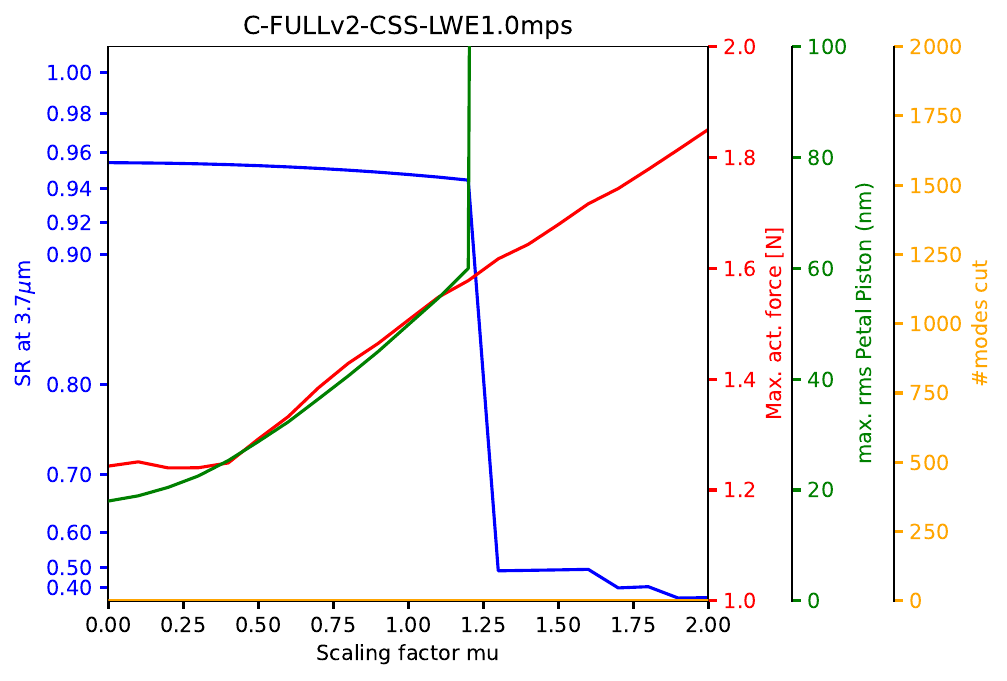}
\includegraphics[width=0.45\textwidth]{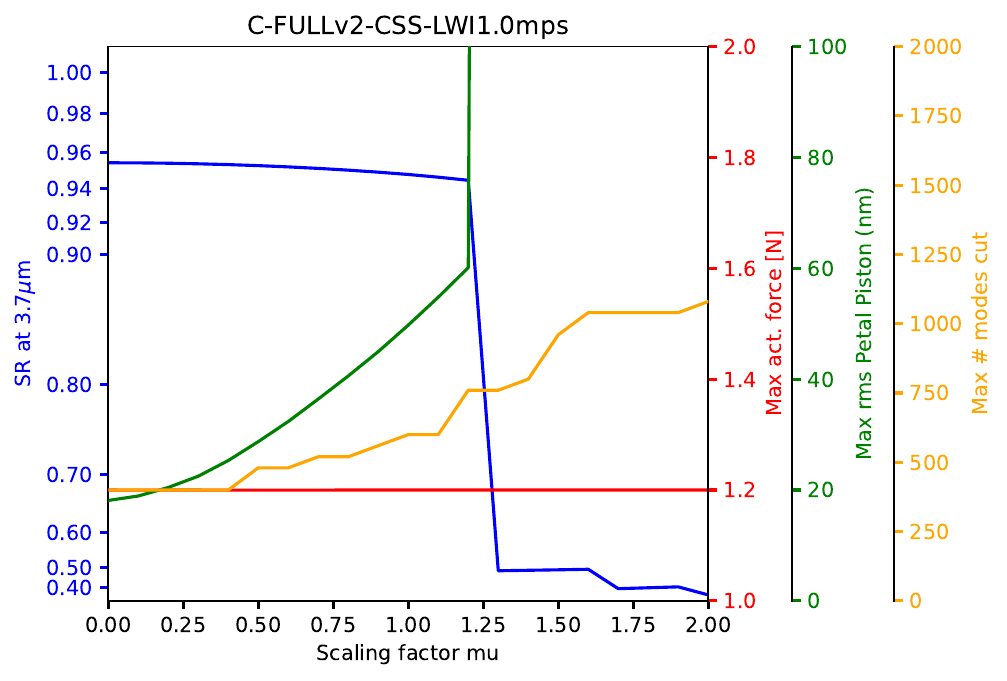}

\end{center}
\caption[]
{\label{fig:C-FULL-CCS}
Performance and CCS response for different scalings of the 1.0\,m/s low-wind case. The blue curve denotes the Strehl ratio, red denotes the maximum force required by any actuator at any time during the simulation, green shows the measured petal piston rms as usual, and the orange curve shows the maximum number of modes dropped from the command vector at any time during the simulation. In the simulation underlying the left-hand figure, mode dropping was disabled to show the strain on the maximum actuator force when the impact of the low-wind condition increases.
}
\end{figure}

Listing \ref{lst:ccspars} in Sec.~\ref{subsec:compass_modules} demonstrates the limits set by the CCS in terms of actuator force, gap, and speed. The low-wind scenario is an interesting test case, as the high optical path differences close to the spiders will demand high actuator forces to compensate, and one might expect petalling problems should the CCS limit the forces or start dropping frames.

Fig.\ref{fig:C-FULL-CCS} shows that such expectations are unfounded. Deviating from the usual scheme, we now plot in red the maximum force required by any actuator at any point in time during the simulation, while in orange we plot the maximum number of modes dropped from the command vector at any point in time during the simulation. In the simulation underlying the left-hand plot in Fig.\ref{fig:C-FULL-CCS}, dropping of modes was disabled in order to track the maximum force required for correcting the impact of the low-wind condition.

We start to see an impact on actuator forces from a scaling factor of $\sim$0.4, with the required maximum force increasing linearly with the scaling factor as expected. At the nominal amplitude of the low-wind impact, the required maximum force is about 1.5\,N. On the right-hand plot of Fig.~\ref{fig:C-FULL-CCS}, mode dropping is enabled, and so the required maximum force never exceeds the threshold of 1.2\,N. We see that even without the low-wind condition, there are occasions when up to $\sim$400 modes get dropped from the command vector. Note that a detailed analysis shows that these dropping events only occur about once every couple of seconds. Yet, the number of dropped modes increases with the scaling factor, as expected, also due to the increasing required maximum force.

Note that while the sharp drop in performance for a scaling factor of 1.25 coincides with the number of dropped modes jumping above 700, this is not the cause for the performance breakdown. In other experiments, we have actually seen up to 2000 modes dropped without a significant impact on performance (remember, it does not happen all the time, it usually only affects individual frames several seconds apart), and in these particular low-wind circumstances, we see petalling problems evolve and quickly escalate to crash the loop even when the CCS is not enforcing any limits.

\subsection{Super Hero}

The various robustness analyses conducted generally started from the baseline configuration and introduced a single adverse effect to examine how robust the system reacts concerning this very effect. In reality, an AO system will always be subject to a combination of many, if not all, of these effects, plus Murphy's law.

In order to anticipate the true performance of our system, we have combined a number of effects and introduced several "known unknown unknowns," such as modeling errors when generating the synthetic command matrices (CMs).

Tab.~\ref{tab:super_hero} summarizes the errors we have introduced into the system {\it after} calibrating for nominal conditions. That means we simulated the case where we have errors that we either do not know about during the generation of the command matrix (CM), or we don't do anything about them. In terms of the latter, we are running with full wind-induced tip-tilt, a total of 21 missing M1 segments arranged in 3 flowers, and the M4 influence functions (IFs) differ between the model used during calibration and that in closed-loop operation.

In addition, we introduce a mis-rotation and mis-registration of M4 with respect to the model used during CM generation. We also assume that the modulator is actually operating at a different amplitude than we thought, we have 130\,nm of NCPA, and an initial petalling error on a single petal of 250\,nm.

\begin{figure}
\begin{center}
\includegraphics[width=0.45\textwidth]{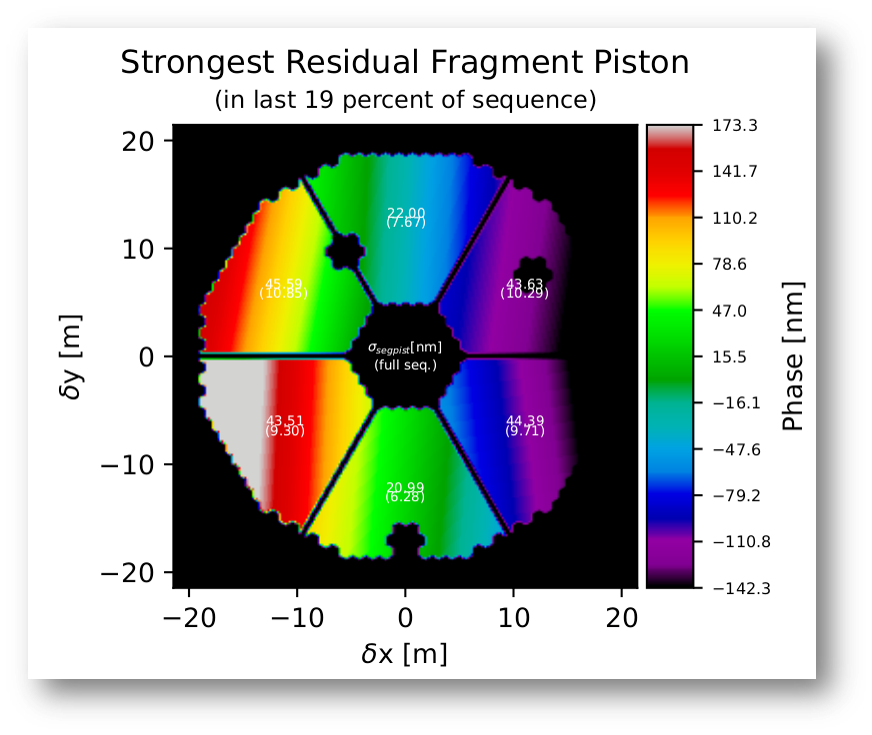}
\includegraphics[width=0.45\textwidth]{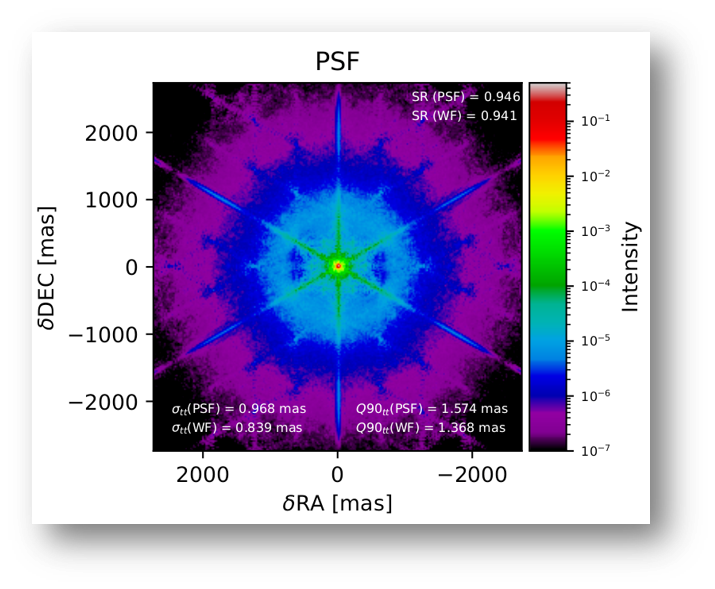}
\end{center}
\caption{\label{fig:superhero_performance}Performance summary graphs for operation under "super hero" conditions. The Strehl ratio remains at a stunning 94.6\%! }
\end{figure}

Fig.~\ref{fig:superhero_performance} summarizes the result. All the introduced errors combined add a total of about 40\,nm rms wavefront error, making the Strehl number at 3.7\,$\mu$m drop to 94.6\%! The residual tip-tilt image motion remains sub-milliarcsecond, and the tip-tilt removed residual piston errors have an rms lower than 10\,nm! In summary, the system is outstandingly robust not only against individual atrocities thrown at it but also against a real-world combination of them!

\begin{table}[hb]
\begin{center}
\caption[Super Hero Configuration]{Errors terms introduced at amplitudes assumed to represent residuals after compensation by various techniques. These were all introduced for closed-loop operation, {\it after} calibration for nominal conditions.}
\label{tab:super_hero}
\begin{tabular}{|ll|ll|} \hline
\multicolumn{2}{|c|}{Configuration} & \multicolumn{2}{|c|}{Scaled Errors} \\
\hline
Controller & leaky PI & Mis-rotation & 0.2$^\circ$ \\
Saturation management & $F<1.2$\,N, $v<2.5$\,m/s, & Mis-registration & 0.2 subap (0.1\,m) \\
& $\Delta<30$\,$\mu$m & Mis-modulation & 0.2\,$\lambda/D$ \\
Wind-induced errors & ON (TT, telescope, etc.) & Initial petal error & 250\,nm \\
NCPA & 130\,nm & Pupil offset by pyr. prism angles & 0.14\,pix \\
Missing segments & 3 flowers && \\
IF model error & YES &&\\
\hline
\end{tabular}
\end{center}
\end{table}

\section{MOVING ON - FROM SIMULATION TO S/HRTC}

Now that FDR is passed and the system has been fully simulated up to the super hero configuration, we are moving to replace the simulated soft and hard real-time computer (SRTC and HRTC) functionalities with the actual soft- and hardware. In an initial stage, ongoing now in Q3 2023, the COMPASS environment will remain to simulate the optical propagation through the atmosphere and telescope, passing generated WFS data to the actual HRTC and receiving back the command vector to apply in the simulated CCS and DMs. The CM generation, telemetry functionalities, etc., will be performed on the actual SRTC.

In a later stage, our reconstruction and control scheme will be verified at the large binocular telescope (LBT), in a laboratory setup, and eventually in the METIS instrument itself, first coupled to a telescope simulator and finally to the ELT.


\begin{figure}
\begin{center}
\includegraphics[width=0.7\textwidth]{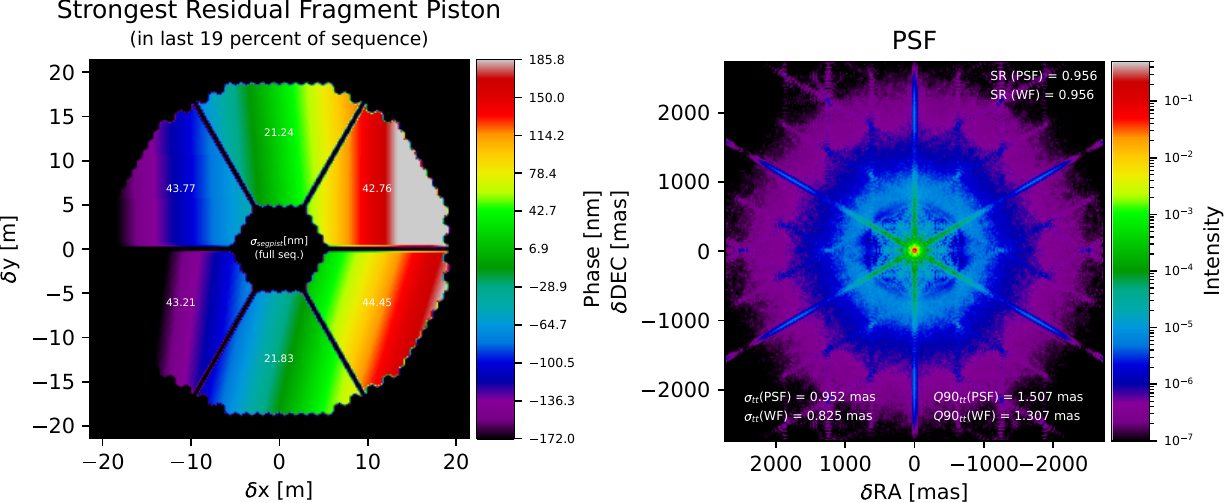}
\end{center}
\caption[]
{\label{fig:hrtc_performance}
Performance summary graphs for operation with the actual HRTC in the loop. The Strehl ratio  improves to 95.6\%, implying a wavefront error rms diminished by $\sim 20$\,nm!
}
\end{figure}

The first step, the inclusion of the HRTC into the simulation pipeline, has been completed in mid-July 2023. A detailed description of the METIS RTC system can be found in \cite{bertram23}. The performance achieved is actually slightly {\it better} than in pure simulation, with the overall wavefront error rms being about 20\,nm lower than in the standard setup. Here we actually suspect a new COMPASS version to be the root cause, which had to be introduced in order to operate in the more modern CUDA environment required by HRTC libraries. The graphical summary report of the run is shown in Fig.~\ref{fig:hrtc_performance}.

\section{CONCLUSION}

In summary, METIS has completed the FDR phase, and SCAO simulations have shown that whatever we have thrown at the system, we can achieve a stable performance of about 95.5\% Strehl at 3.7\,$\mu$m under the most adverse circumstances. A notable exception is wind speeds of or below 0.5\,m/s in the telescope pupil. Even in combination, saturation management, wind-shake, missing segments, modeling errors, NCPAs, and initial petalling only have a minor impact on our system performance.

Overall, the results indicate that the METIS SCAO system is robust to various challenges, including guide-star magnitude degradation, NCPAs up to 300\,nm, and moderate mis-registration and mis-rotation. However, the low-wind effect can significantly impact the system's performance, and further mitigation measures may be needed to maintain high contrast under such conditions.

Since FDR, we have moved on and plugged the HRTC prototype into our simulations, which now performs its foreseen task of turning WFS images into the wavefront command vector. The performance remains perfectly nominal, as is to be expected. The next step is to also move the SRTC tasks to the corresponding prototype.

\acknowledgments 

The authors would like to thank the organizers and all participants of the AO4ELT conference for the wonderful time in Avignon, excellent presentations, endless stimulating discussions, and the specially arranged sunset across the {\it pont d'Avignon}.

Special thanks to Silvia Scheithauer for proof-reading and thus improving the manuscript, and ensuring the non-violation of any conceivable export regulation.

This article was spell- and stylechecked, but not generated by chatGPT 4.0.

\printbibliography 
\end{document}